\documentclass[aps,prl,twocolumn,superscriptaddress,groupedaddress]{revtex4} 
\pdfoutput=1
\usepackage[version=3]{mhchem} 
\usepackage[utf8]{inputenc}
\usepackage[dvipsnames]{xcolor}
\usepackage{pdfpages}
\usepackage[T1]{fontenc}
\usepackage{soul}
\DeclareMathOperator{\atan2}{atan2}

\definecolor{FF}{RGB}{14, 159, 14}
\definecolor{mypink2}{RGB}{219, 48, 122} 
\begin{document}

\title{Level attraction and idler resonance in a strongly driven Josephson cavity}


\author{F. Fani Sani}
\email{fatemeh.fanisani@uwaterloo.ca}
\affiliation{Kavli institute of Nanoscience, Delft University of Technology, PO Box 5046, 2600 GA Delft}
\affiliation{Institute for Quantum Computing, University of Waterloo, Waterloo, Ontario N2L 3G1, Canada}

\author{I.~C. Rodrigues}
\affiliation{Kavli institute of Nanoscience, Delft University of Technology, PO Box 5046, 2600 GA Delft}

\author{D. Bothner}
\affiliation{Kavli institute of Nanoscience, Delft University of Technology, PO Box 5046, 2600 GA Delft}
\affiliation{Physikalisches Institut and Center for Quantum Science in LISA$^+$, Universit\"at T\"ubingen, 72076 T\"ubingen, Germany}

\author{G.~A. Steele}
\affiliation{Kavli institute of Nanoscience, Delft University of Technology, PO Box 5046, 2600 GA Delft}

\begin{abstract}

%
%

Nonlinear Josephson circuits play a crucial role in the growing landscape of quantum information and technologies. 
%
The typical circuits studied in this field consist of qubits, whose anharmonicity is much larger than their linewidth, and also of parametric amplifiers, which are engineered with linewidths of tens of MHz or more. 
The regime of small anharmonicity but also narrow linewidth, corresponding to the dynamics of a high-$Q$ Duffing oscillator, has not been extensively explored using Josephson cavities. 
Here, we use two-tone spectroscopy to study the susceptibility of a strongly driven high-$Q$ Josephson microwave cavity. 
Under blue-detuned driving, we observe a shift of the cavity susceptibility, analogous to the AC Stark effect in atomic physics.  
When applying a strong red-detuned drive, we observe the appearance of an additional idler mode above the  bifurcation threshold with net external gain. 
Strong driving of the circuit leads to the appearance of two exceptional points and a level attraction between the quasi-modes of the driven cavity. 
%
Our results provide insights on the physics of driven nonlinear Josephson resonators and form a starting point for exploring topological physics in strongly-driven Kerr oscillators.

\end{abstract}

\maketitle

\section*{Introduction }

Circuit quantum electrodynamics (cQED) \cite{Blais21}, based on nonlinear superconducting circuits, has laid the foundation for the current tremendous growth of the superconducting quantum technology field. Microwave circuits inheriting a large anharmonicity from the nonlinear nature of Josephson junctions have been used to engineer artificial atoms \cite{Krantz19}, for Fock and cat state generation \cite{Hofheinz08, Kirchmair13, Leghtas15}, for synthesizing arbitrary quantum states of light \cite{Hofheinz09}, to demonstrate the lasing effect from a superconducting qubit \cite{Astafiev07} and for engineering single photon sources \cite{Zhou20}. 
Josephson circuits where the nonlinearity does not exceed the cavity decay rate have been equally successful. These have been utilized as a platform to investigate the dynamical Casimir effect \cite{Wilson11, Lahteenmaki13}, as systems to perform dispersive readout of superconducting qubits \cite{Lin14,Krantz16} and have been recently used for the realization flux-mediated optomechanical systems \cite{Rodrigues19,Zoepfl20,Schmidt20,Bothner21} and photon-pressure systems \cite{Bothner21PP, Eichler18}.
Also in the context of small nonlinearities, a common application is the engineering of Josephson Parametric Amplifiers (JPAs) \cite{Yamamoto08, Eom12, Yaakobi13, Zhou14, Castellanos07, Winkel20,Gross2017}. 
Such systems have approached quantum-limited amplification \cite{Bergeal10,Macklin15} and allowed for squeezing of vacuum fluctuations \cite{Castellanos08,Zhong13}. 

Typically the nonlinear circuits utilized for the realization of JPAs are designed to be highly overcoupled and with a large decay rate in order to optimize their gain bandwidth ratio.
However, there has been a growing interest lately in strongly driven  high-quality-factor (high-$Q$) Josephson circuits. For example, they have recently been used to cool a nanobeam with a blue-detuned pump in flux-mediated optomechanical systems \cite{Bothner21}, to study the quantum nature of nonlinear damping in microwave circuits \cite{Gely21} or for high-sensitivity current detection by upconversion \cite{Felix20}. Furthermore, circuits with a flux-tunable nonlinearity have also been used to explore the crossover between a classical Duffing oscillator and a Kerr parametric oscillator \cite{Yamaji20, Andersen20}.

Bringing a nonlinear oscillator far beyond its bifurcation point is a regime which has been explored in the context of driven micromechanical resonators \cite{Stambaugh06}, optical systems \cite{Drummond80, Khandekar15}, and has been recently used for the detection of squeezing of a mechanical mode \cite{Huber20, Ochs21}. 
Past its bifurcation threshold the system enters a regime of bistability where two stable solutions exist \cite{Dykman80, Marthaler06}. In superconducting circuits the switching between these bistable branches has been explored close to bifurcation for the engineering of the Josephson bifurcation amplifier \cite{Siddiqi04, Vijay09}, a technology which was later on utilized for fast measurement and single-shot read-out of superconducting qubits \cite{Mallet09}.
Going further beyond the bifurcation limit, however, is a regime which is typically unexplored within JPAs as this suppresses the maximum achievable gain of an amplifier \cite{Planat19}. Nevertheless, superconducting circuits could be a powerful platform to explore this regime as they grant a large design flexibility and can be operated in well controlled environments.

Here we explore the physics behind a weakly nonlinear high-$Q$ Josephson circuit which is driven far beyond its bifurcation point. We study the two-tone response of the resonator for a pump tone placed above or below the cavity resonance and investigate the modification to its driven susceptibility by extracting the pump-modified resonance frequency and decay rates seen by the probe. While a blue detuned pump gives rise to the well known AC stark shift \cite{Delone99}, a red detuned pump will bring the system above its bifurcation threshold and place the oscillator in a bistable operation regime. In addition to a strong modification of the cavity mode susceptibility, which we later define as signal mode, a second idler mode exhibiting a net output gain will emerge upon probing the system with a weak tone. 
After investigating the characteristics of these two new modes, we focus on the dependence of the driven cavity state on intracavity photon number and uncover the existence of level attraction between the signal and idler modes. Our results provide new insights on the understanding of strongly driven high-$Q$ Kerr oscillators and reveal the potential for operating these systems well above their bifurcation threshold.

\section*{Results}

\begin{figure}[h!]
\centerline{\includegraphics[trim = {0.0cm, 1.0cm, 1.0cm, 0.0cm}, clip=True,width=0.55\textwidth]{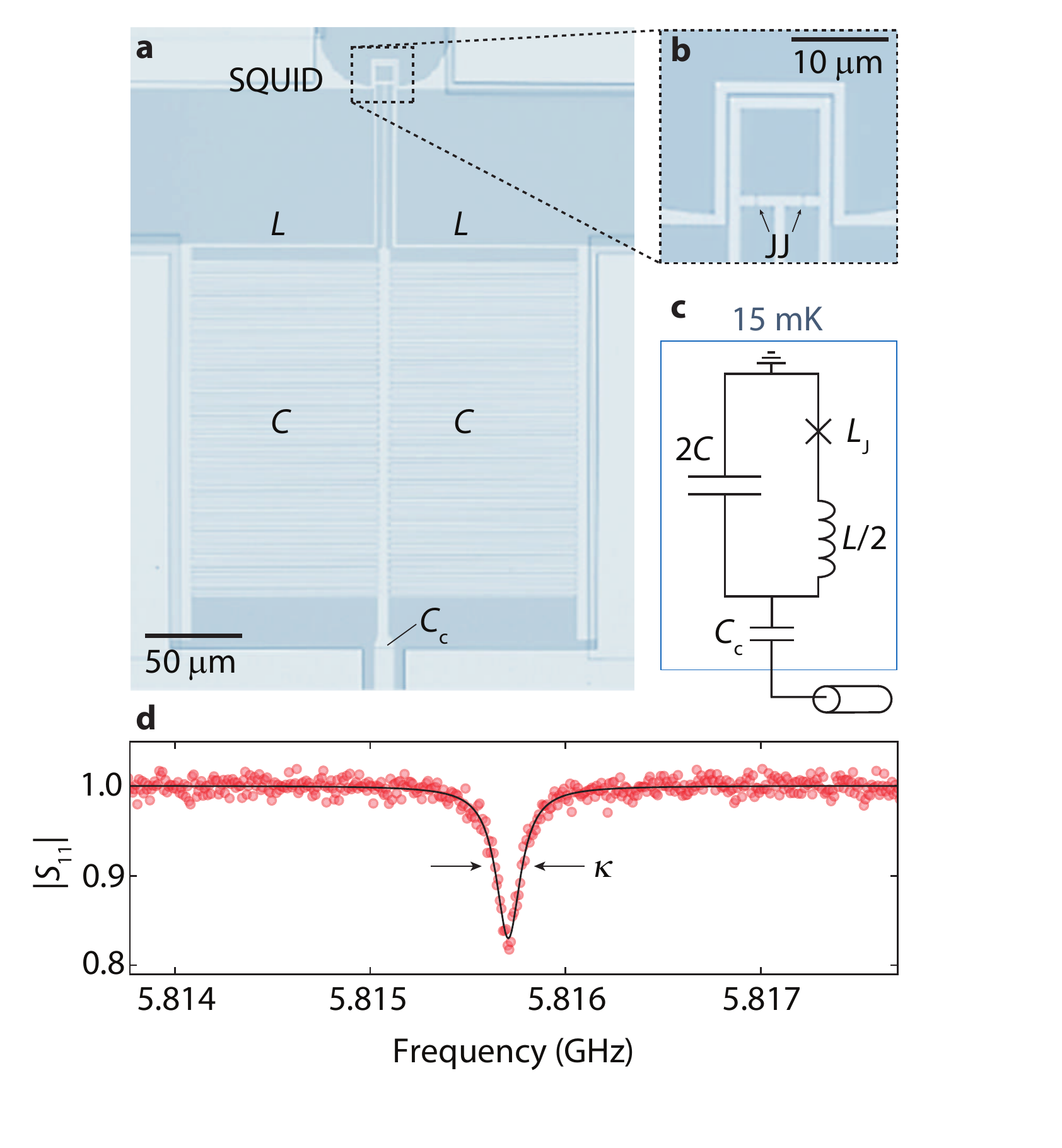}}
\caption{\textbf{A lumped element Josephson cavity}.
Panel \textbf{a} shows a micrograph of the full device and corresponding labeling of the main circuit elements. The coupling capacitor is $C_\textrm{c} = 2$~fF, each of the interdigitated capacitors (IDC) is given by $C = 0.5$~pF and the linear inductance contribution of the circuit by two wires that link the IDCs and the SQUID as $L/2 = 742$~pH. The SQUID is visible at the top part of the image, and a zoom-in in which the two nanobridge Josephson junctions (JJ) are shown, is presented in \textbf{b}. Panel \textbf{c} shows a schematic representation of the circuit model where $L_\textrm{J} = 20$~pH is the SQUID inductance. In panel \textbf{d} the measured reflection magnitude $|S_\textrm{11}|$ and its corresponding fit are shown as red circles and black line, respectively. From the fit parameters we extract a resonance frequency of $\omega_0 = 2\pi \cdot 5.8157$ GHz, a total decay rate $\kappa = 2\pi\cdot157$~kHz and an external decay rate $\kappa_\textrm{e}=2\pi\cdot12.5$~kHz.}
\label{fig1:sems}
\end{figure}

The device we study here consists of a microwave Josephson cavity capacitively coupled to a coplanar waveguide feedline by means of a coupling capacitor $C_\textrm{c}$, and containing a Superconducting QUantum Interference Device (SQUID). Optical images of the full device and a zoom-in of the $10~\mu\textrm{m}\times10~\mu\textrm{m}$ SQUID including its two nanobridge Josephson junctions are shown in Figs.~\ref{fig1:sems}\textbf{a} and \ref{fig1:sems}\textbf{b}, respectively.
%
When an external magnetic field is applied to the device, the Josephson inductance of the SQUID and therefore the cavity resonance frequency can be tuned. 
In the absence of an applied magnetic flux in the loop, i.e. at the SQUID cavity sweet-spot, the circuit model of the device can be simplified to a LC resonator, where most of its capacitance $2C$ comes from two symmetric interdigitated capacitors (IDCs) and its total inductance $\frac{L}{2}+ L_\textrm{J}$ arises from a combination of two inductor wires of inductance $L$ and the sweet-spot Josephson inductance $L_\textrm{J}$ representing the SQUID. In the work reported here the cavity will always be operated at its sweet-spot. A simplified schematic of the circuit is shown in Fig.~\ref{fig1:sems}\textbf{c}, once the device is mounted in a dilution refrigerator with a base temperature of approximately $T_\textrm{b} \approx 15 \, \textrm{mK}$.

When probing the system with a weak tone, we observe an absorption dip in the reflected signal, as shown by the cavity response $|S_{11}|$ in Fig.~\ref{fig1:sems}\textbf{d}. From a fit we extract a resonance frequency $\omega_0=2\pi \cdot 5.8157$ GHz, a total decay rate $\kappa=2\pi\cdot157$ kHz, an external decay rate $\kappa_\textrm{e}=2\pi\cdot12.5$ kHz and an internal decay rate $\kappa_\textrm{i}=2\pi\cdot144.5$ kHz. Furthermore, the ratio between internal and external decay rates $\kappa_\textrm{e}/\kappa_\textrm{i} \sim 0.1$ tells us that the power is mostly dissipated inside the cavity, what is usually considered as an undercoupled detection scheme ($\kappa_\textrm{e}/\kappa_\textrm{i} < 1$).\\


\begin{figure*}[htb]
\begin{center}
\includegraphics[trim = {1.0cm, 6.0cm, 4.0cm, 1.0cm}, width=1\linewidth]{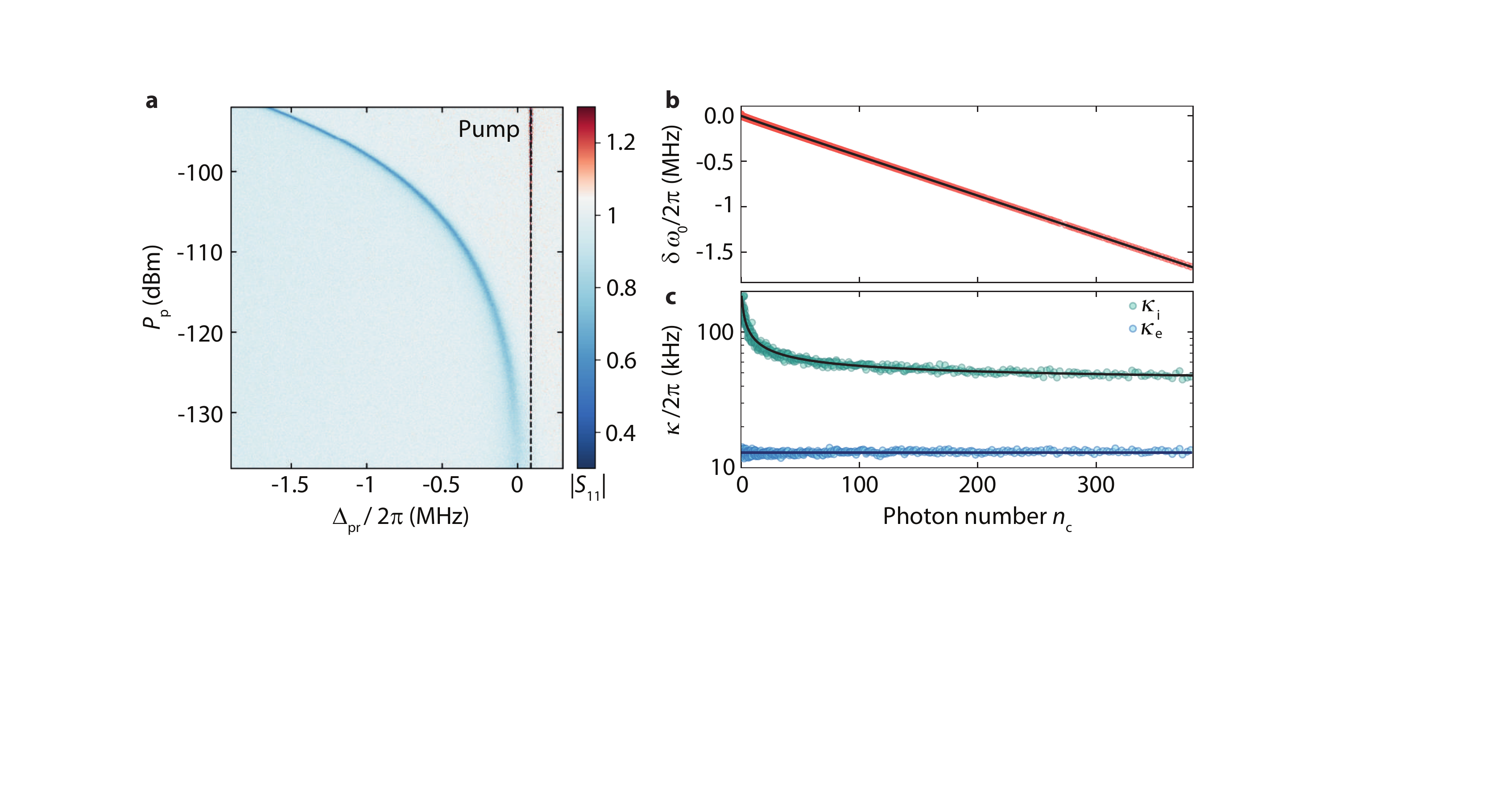}
\caption{\textbf{Observation of the AC stark shift and TLS losses of a Josephson cavity using two-tone spectroscopy.}
In panel \textbf{a} the normalized cavity response $|S_\textrm{11}|$ is shown in a colormap as a function of on-chip pump power $P_\textrm{p}$ and probe detuning from the bare cavity resonance $\Delta_\textrm{pr} = \omega_\textrm{pr} - \omega_0$, for a pump tone sitting at the point indicated by the dashed black line with detuning $\Delta_\textrm{p}/2\pi = 80$ kHz. Here the pump power was increased in steps of $0.1$~dBm. For each horizontal line of the colormap presented in \textbf{a} we fit the response spectrum with Eq.~(\ref{eq:linearS11}) and extract parameters such as dressed resonance frequency $\omega_\textrm{s}$, and internal and external linewidths, $\kappa_\textrm{i}$ and $\kappa_\textrm{e}$. Additionally, we estimate the intracavity photon number $n_\textrm{c}$ from a second fit of the cavity response based on the theoretical model of a driven Kerr cavity (see Supplementary Note 3). Panel \textbf{b} shows the extracted resonance frequency shift $\delta\omega_\textrm{0} = \omega_\textrm{s} - \omega_0$ versus $n_\textrm{c}$. Together with the data we plot the theoretically predicted resonance frequency shift (black line) for an anharmonicity $\mathcal{K}/2\pi = -2.5$~kHz, which arises from the nonlinear inductance of the circuit. This is consistent with the measured pump line attenuation of $\sim107$~dB. In panel \textbf{c} we plot the extracted decay rates as function of $n_\textrm{c}$. While the external linewidth $\kappa_\textrm{e}$ (blue circles) stays approximately constant over the presented range ($\kappa_\textrm{e} \approx 2\pi \cdot 13$ kHz), the internal loss rate $\kappa_\textrm{i}$ (teal circles) shows a considerable decrease with photon number, reaching a value of $2\pi \cdot 50$ kHz. Such behavior is well described by the saturation of dielectric losses from two-level systems (TLSs), as shown by the fit curve represented by the black line.}
\label{fig2:blue_detune}
\end{center}
\end{figure*}

In Fig.~\ref{fig2:blue_detune} we present data exploring the cavity response under the excitation from two microwave tones: a pump tone $\omega_\textrm{p}$ which strongly drives the circuit and a weak probe tone $\omega_\textrm{pr}$ that has low enough power to not drive, but only to measure the response of the device. Fig.~\ref{fig2:blue_detune}\textbf{a} shows a colormap of the reflection magnitude $|S_\textrm{11}|$ versus probe detuning $\Delta_\textrm{pr} = \omega_\textrm{pr} - \omega_0$ and on-chip pump power $P_\textrm{p}$ for a blue-detuned pump. The pump frequency $\omega_\textrm{p}$ is chosen slightly above the cavity resonance $\omega_0$ with detuning $\Delta_\textrm{p}= \omega_\textrm{p} - \omega_0 < \kappa$, as indicated by the dashed black line. 


The first and most obvious experimental observation is the gradual resonance frequency shift towards lower values as the pump power is increased. The origin of this shift emerges from the Josephson nonlinearity of the cavity: increasing the pump power raises the intracavity photon number $n_\textrm{c}$ and results in a dressed resonance frequency $\omega_\textrm{s}$ that is subsequently detected by the weak probe signal, analogously to the AC Stark shift in atomic physics \cite{Delone99}.  
To quantify this, we fit the cavity response $S_{11}$ with the reflection response function of a linear LC circuit 
\begin{equation}
S_{11} = 1 - \kappa_\textrm{e}\chi_\textrm{c}
\label{eq:linearS11}
\end{equation}
with the susceptibility $\chi_\textrm{c}^{-1}(\omega) = \frac{\kappa}{2} + i(\omega-\omega_{s})$, and extract the cavity parameters such as decay rate and resonance frequency. We note that in this regime our system can still be treated as a linear cavity with a modified resonance frequency.
In Fig.~\ref{fig2:blue_detune}\textbf{b}, we plot the extracted resonance frequency shift between the dressed and the bare resonance frequency $\delta\omega_0 = \omega_\textrm{s} - \omega_0$ as a function of the intracavity photon number $n_\textrm{c}$. The latter was estimated from a fit using a theoretical model based on the linearized equation of motion of a Kerr cavity (see Supplementary Note 3). The theoretically predicted frequency shift for a resonator with a Kerr nonlinearity is plotted simultaneously with the experimental data as black line and red circles, respectively. For the theory line we use an anharmonicity value of $\mathcal{K} /2\pi = - 2.5$ kHz, which is consistent with the measured pump line attenuation.
Based on this value, we can model our SQUID cavity as a harmonic oscillator with a weak Kerr nonlinearity $\mathcal{K}\ll\kappa$. Additionally, we can also analytically estimate the device anharmonicity based on the circuit parameters $\mathcal{K} = - \frac{e^2}{4C}\left(\frac{2L_\textrm{J}}{L + 2L_\textrm{J}}\right)^3 = - 2\pi \cdot 380~$Hz, a considerably smaller value compared to the one extracted from the measurements. An explanation for this discrepancy could be an asymmetry between the individual Josephson junctions inductance. 

In addition to the frequency shift, a closer look to Fig.~\ref{fig2:blue_detune}\textbf{a} shows evidence of the cavity resonance getting sharper and deeper as the pump power is increased. This suggests that the internal loss rate of the cavity is decreasing with increasing pump power. To investigate this, we plot the extracted external $\kappa_\textrm{e}$ and internal $\kappa_\textrm{i}$ decay rates versus $n_\textrm{c}$ in Fig.~\ref{fig2:blue_detune}\textbf{c}.
%
\begin{figure*}[htb]
\centering
\includegraphics[trim = {0.0cm, 5.7cm, 5.0cm, 2.0cm}, width=1\linewidth]{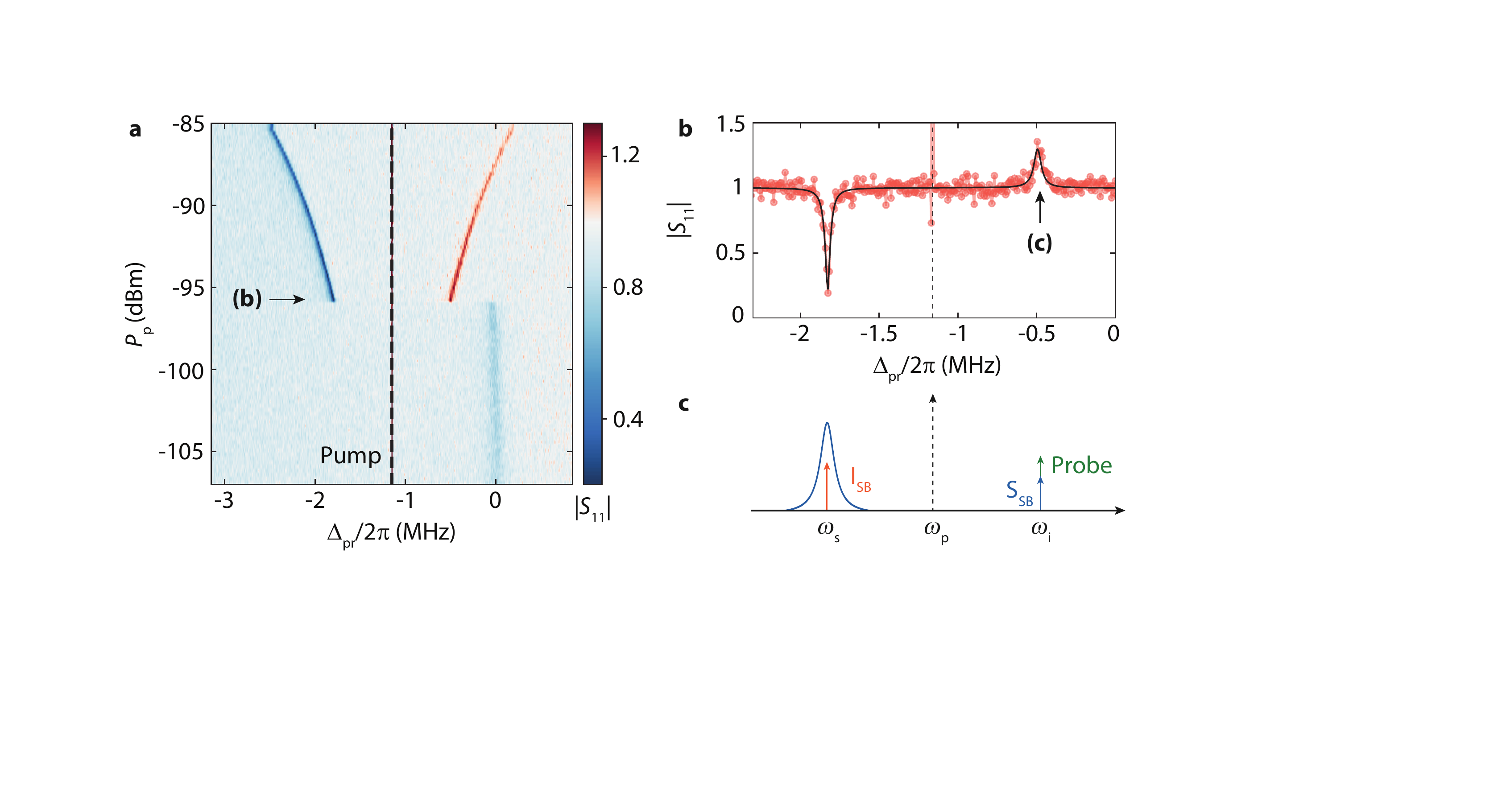}
\caption{\textbf{Detection of an idler mode with net output gain in a regime beyond bifurcation}. Panel \textbf{a} shows a colormap of the cavity response $|S_\textrm{11}|$ as a function of on-chip pump power $P_\textrm{p}$ and probe detuning from the bare resonance $\Delta_\textrm{pr} = \omega_\textrm{pr} - \omega_{0}$. The pump tone is fixed at a constant detuning from the bare resonance $\Delta_\textrm{p}/2\pi =- 1.16$ MHz as indicated by the black dashed line. When the pump power is increased above a certain threshold ($P_\textrm{p} = -97 $ dBm), there is an abrupt jump from a faint dip to a dip-peak region. A linescan above the switching point ($P_\textrm{p} = -96 $ dBm) is shown in \textbf{b} and is indicated by the black arrow. 
Panel \textbf{b} displays a linescan showing the dressed cavity mode and a prominent idler-resonance. Red circles are data and the black line is a theory line used with Eq.~(\ref{eq:S11}). The large amplitude of the pump tone, which is a single frequency point in the experimental data is ignored in the fit routine.  
\textbf{c} A schematic diagram is depicted for a probe tone at $\Delta_\textrm{pr}/2\pi \sim - 0.5$ MHz, indicated by the black arrow in \textbf{b}. The cavity density of states is shown as a blue Lorentzian curve, while the probe and pump tones are represented by the green solid and black dotted arrows, respectively. The idler ($\textrm{I}_\textrm{SB}$) and signal ($\textrm{S}_\textrm{SB}$) sidebands, generated by four-wave mixing, are presented in red and blue arrows. The cavity nonlinearity is responsible for this mixing mechanism and when the signal or the idler sideband falls within the cavity linewidth, it gives rise to parametric amplification.
}
\label{fig3:red_detune}
\end{figure*}
Interestingly, we observe a strongly power dependent internal loss rate. Such an observation is common in superconducting resonators \cite{Pappas11} and can be attributed to the saturation of dielectric losses due to two-level systems (TLSs), which are known to be located at material interfaces and film surfaces \cite{Woods18}. We fit the extracted internal decay rate $\kappa_\textrm{i}$ (teal circles) with a TLS model {$\kappa_\textrm{i} = \kappa_0 +\frac{\kappa_1}{\sqrt{1+{n_\textrm{c}}/{n_\textrm{crit}}}}$}, shown as dark gray line in Fig.~\ref{fig2:blue_detune}\textbf{c}. The model describes well the experimental data for $\kappa_1 = 2\pi \cdot  175$~kHz, $n_\textrm{crit} =  1$ and the saturation value $\kappa_0 = 2\pi \cdot 40$ kHz, showing the significant increase of the cavity internal quality factor when the system is driven by a strong pump.
The external cavity decay rate $\kappa_\textrm{e}$ (blue circles) is found to be approximately constant for all pump powers with an average value of $2\pi \cdot 13$ kHz (navy blue line).\\

An intriguing experimental setting is to investigate the cavity response when a strong pump is placed red-detuned from the bare cavity resonance.
The outcome of this measurement is shown in Fig.~\ref{fig3:red_detune}\textbf{a}, where the reflection magnitude $|S_\textrm{11}|$ is plotted as a function of probe detuning $\Delta_\textrm{pr}$ and on-chip pump power $P_\textrm{p}$ for a red-detuned pump with $\Delta_\textrm{p}/2\pi = -1.16$ MHz.
Whereas for the case of a blue-detuned pump the dressed cavity resonance shifts away from the drive for increasing power, for the red-detuned case the dressed cavity resonance moves towards the pump, which leads to a further increase in intracavity photon number. When the driving power reaches a specific threshold, the cavity hits its bifurcation point and the system subsequently enters a bistable regime with two possible solutions in response to the pump field: a high-amplitude and a low-amplitude branch, in which the currents oscillating in the circuit at the pump frequency have either a low or high amplitude. Crossing the bifurcation threshold allows the cavity to operate in either of these regimes and to even punctually switch between them, as visible by the abrupt jump of the cavity to the other side of the pump around $P_\textrm{p} = -97$~dBm (cf. Fig.~\ref{fig3:red_detune}\textbf{a}). This indicates that the cavity is afterwards operated on the high-amplitude branch and from there onwards, the dressed cavity response shifts similarly to the case of a blue-detuned pump. Note that although the cavity is in a nonlinear bifurcated state in response to the strong pump, the susceptibility seen by the weak probe is still that of a linear response, with a Lorentzian lineshape that is downshifted in frequency compared to the unpumped circuit.

The most peculiar outcome of crossing the cavity bifurcation point is the appearance of an additional mode that has a net output gain at a frequency symmetric to the dressed cavity resonance with respect to the drive. This is clearly displayed in Fig.~\ref{fig3:red_detune}\textbf{b}, which shows a linescan of $|S_\textrm{11}|$ straight after the switching point, for an on-chip pump power of $-96$ dBm.
To understand the physics behind the appearance of this feature, we additionally illustrate the intracavity fields in Fig.~\ref{fig3:red_detune}\textbf{c}, for the specific case of a probe tone frequency indicated by the label \textbf{(c)} in Fig.~\ref{fig3:red_detune}\textbf{b}. Here, the pump and probe intracavity fields, which are generated by external input signals, are shown in black dotted and green solid arrows, respectively. In this driving configuration, the nonlinearity of the cavity generates two additional intracavity fields (red and blue arrows) via a four-wave mixing process. A field with the same frequency as the input probe tone, commonly referred as the signal sideband ($\textrm{S}_\textrm{SB}$) and the field arising symmetrically on the other side of the pump tone, commonly referred to as the idler sideband ($\textrm{I}_\textrm{SB}$) in the nomenclature of parametric amplifiers. As the amplitude of the internally generated fields depends on the driven Kerr cavity susceptibility
\begin{equation}
\chi_\mathrm{g} = \frac{{\chi}_\mathrm{p} (0) }{1 - \mathcal{K}^2n_\mathrm{c}^2{\chi}_\mathrm{p}(0){\chi}_\mathrm{p}^*(2\Omega_\textrm{pp})},
\end{equation}
with ${\chi}_\mathrm{p}^{-1}(\Omega) = \frac{\kappa}{2} + i\left(\Delta_\textrm{pr}  - 2\mathcal{K}n_\mathrm{c} + \Omega\right)$ and $\Omega_\textrm{pp} = \omega_\textrm{p} - \omega_\textrm{pr}$, when the probe scans through the opposite side of the dressed cavity mode, it will also detect an additional lineshape, which we refer to as idler-resonance. This happens as the probe field constructively interferes with the signal sideband generated by four-wave mixing. In fact, the appearance of an idler-resonance should also happen below the bifurcation threshold. However, the low intracavity photon number in this regime suppresses the signal sideband field amplitude below the noise level, and therefore the mode remains experimentally undetected.

 %
\begin{figure*}[htb]
\centering
\includegraphics[trim = {1.0cm, 6.0cm, 0.0cm, 3.0cm}, width=1\linewidth]{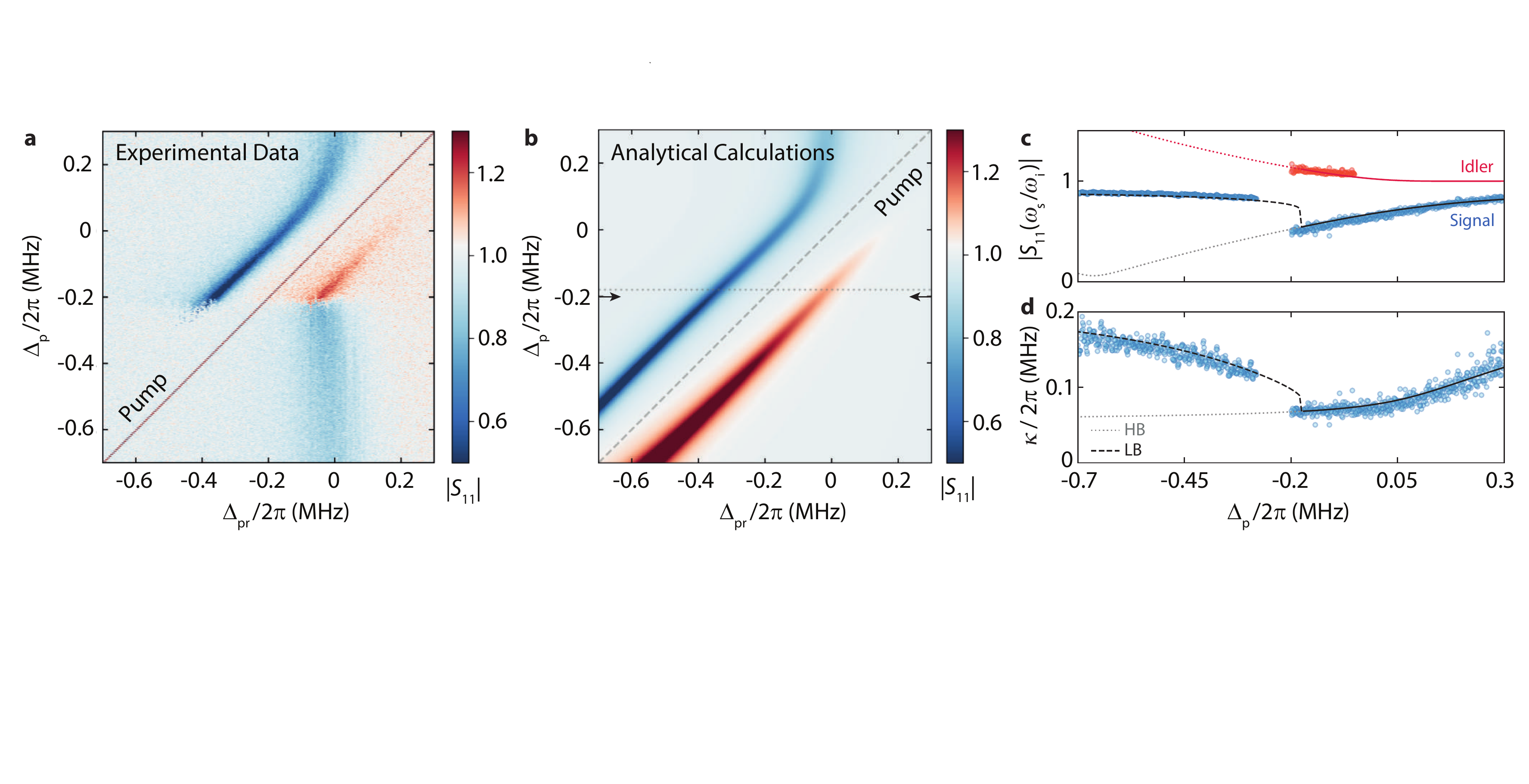}
\caption{\textbf{Frequency locking of a Kerr cavity to a strong parametric drive.}
\textbf{a} The cavity response $|S_\textrm{11}|$ as a function of pump detuning $\Delta_\textrm{p}$ and probe detuning $\Delta_\textrm{pr}$ for a on-chip pump power of $-127$ dBm. The frequency of the pump tone was reduced in steps of $5$~kHz.
Panel \textbf{b} shows a corresponding colormap of the cavity response computed with Eq.~(\ref{eq:S11}). The device parameters were kept constant as extracted from the analysis of Figs.~\ref{fig1:sems} and \ref{fig2:blue_detune}: Line attenuation of $\sim 107$~dB, $\kappa_\textrm{e} / 2\pi \sim 13$~kHz, $\mathcal{K}/ 2\pi = -2.5$ kHz and the power-dependent internal loss rate $\kappa_\textrm{i}$. In \textbf{c} we plot the magnitude of the experimentally detected $|S_{11}|$ on resonance with the signal (blue circles) and idler (red circles) modes, and in \textbf{d} we plot the signal linewidth versus $\Delta_\textrm{p}$. The experimental values were obtained by fitting the full spectrum with a reflection response function that contains two individual modes in the region where the idler is detectable (see Supplementary Note 2). The dotted lines present the analytically estimated values for the high-amplitude branch (gray for signal and red for idler). The dashed black line corresponds to the signal mode of the low-amplitude branch and the full lines are used in the region where there is only a single solution (gray for signal and red for idler). In panel \textbf{c}, the magnitude of idler mode for the low-amplitude branch has been omitted for clarity. We point out that we find a higher level of agreement between the experimental data and the theory by reducing the parameter $\kappa_\textrm{1}/2\pi$ by $15$~kHz.
}
\label{fig4:wt_sim}
\end{figure*}

As evident from the cavity reflection coefficient greater than unity, the signature of an idler resonance is accompanied by amplification of the intput field. This is visible in the colormap shown in Fig.~\ref{fig3:red_detune}\textbf{a} and emphasized in the linecut presented in Fig.~\ref{fig3:red_detune}\textbf{b}, where the black line shows the magnitude of the theoretically predicted response
\begin{equation}
S_\textrm{11} = 1 - \kappa_\textrm{e}\chi_\mathrm{g}.
\label{eq:S11}
\end{equation}
This amplification is common in JPAs, where it is typically occurring within the cavity lineshape. In our case, however, as the cavity is highly undercoupled, we do not observe output field amplification when the signal is resonant with the cavity. Instead we see a deeper absorption dip. This occurs as the intracavity field amplification effectively brings the system closer to critically coupled, leading to a deeper dip in the response spectrum $|S_\textrm{11}|$. Observing amplification of the output field in the signal resonance condition would require a sufficiently large intracavity amplification overcome all internal losses in the device (see Supplementary Note 6 for an example of such data). 
%
This is different, however, for the case of the idler resonant condition where we detect a net output gain, as show in Fig.~\ref{fig3:red_detune}\textbf{b}. Here the input signal is not resonant with the cavity and therefore the interference between the probe and amplified fields is not affected by the cavity resonance. The frequency matching between the idler sideband and the cavity resonance, however, provides a high density of states for the idler photons, a necessary condition for parametric amplification, resulting in the detection of output field amplification even when the input signal is detuned by many linewidths outside of the cavity resonance.\\


Although until now we have only explored the impact of driving power on the cavity response, we can also investigate the phenomenon of idler-resonance by playing with another knob, which is sweeping the pump frequency for a fixed power. In Fig.~\ref{fig4:wt_sim}\textbf{a} the cavity response $|S_\textrm{11}|$ is shown as a function of pump and probe detuning from the bare cavity resonance frequency. The on-chip pump power is chosen to be sufficiently large ($P_\textrm{p} = -127$~dBm) to bring the cavity across its bifurcation point. 

As we sweep the pump frequency from higher to lower values and approach the bare cavity resonance, the system experiences a locking mechanism to the drive. This is visible by the frequency dragging of the signal mode, i.e. the dressed cavity resonance, with decreasing pump frequency. Moreover, as the idler mode always emerges mirrored with respect to the drive tone, this one will also experience a similar dragging as the dressed cavity. Once again, while the cavity is identified by a dip in the spectrum (blue features in Fig.~\ref{fig4:wt_sim}), the idler is visible as a peak above the background level (red features in Fig.~\ref{fig4:wt_sim}). Simultaneously with the locking, also the intracavity pump field becomes larger as the drive comes nearer the bare cavity lineshape, and continues to increase as the pump is detuned further down towards negative values. 
The increase in drive photons arising from attempting to reduce the detuning between the pump and the dressed cavity mode will enhance the signal and idler sidebands, thereby intensifying the signature of both modes upon a probe reflection measurement and achieving higher intracavity field amplification. 

Similarly to what was observed in Fig.~\ref{fig3:red_detune}, in Fig.~\ref{fig4:wt_sim}\textbf{a} we also detect a sharp modification of the cavity response spectrum for a pump detuning around $\Delta_\textrm{p}/2\pi = -0.38$~MHz, which in this case reveals a transition from the high-amplitude to the low-amplitude branch solution. Once the cavity has switched branches and is no longer operated in a metastable regime, the intracavity field amplification is reduced and the idler mode becomes experimentally undetectable. We note that in contrast with the data presented in Fig.~\ref{fig3:red_detune}, here we observe some fast switching between two branches around the transition point which is most probably triggered by noise.

Fig.~\ref{fig4:wt_sim}\textbf{b} shows a colormap of the response based on the analytical model described in Supplementary Note 3, i.e. on Eq.~(\ref{eq:S11}). A clear difference between the data and the theoretical model is the fact that the switching point is predicted to occur for a pump detuning below the range of our measurement. In fact, while in the experiment the cavity switches from a high-amplitude to a low-amplitude branch, the theoretically predicted transition only happens when the system reaches the regime of a single-solution. We note though, that the branch where the system ends up on is highly sensitive to the exact history of the system and to the noise in the device. 
Therefore, the experimental transition point can happen at any frequency when the system is operated above its bifurcation point. In spite of this somewhat expected mismatch, the analytical model shows an excellent qualitative agreement with the experiment as shown in Fig.~\ref{fig4:wt_sim}\textbf{b}. Here the horizontal dashed gray line represent the bifurcation point, which would also be the transition frequency that the cavity would experience if it had been operated in the low-amplitude branch. 

\begin{figure}[h!]
\centering
\includegraphics[trim = {2.0cm, 1,5cm, 2.5cm, 0cm}, width=0.8\linewidth]{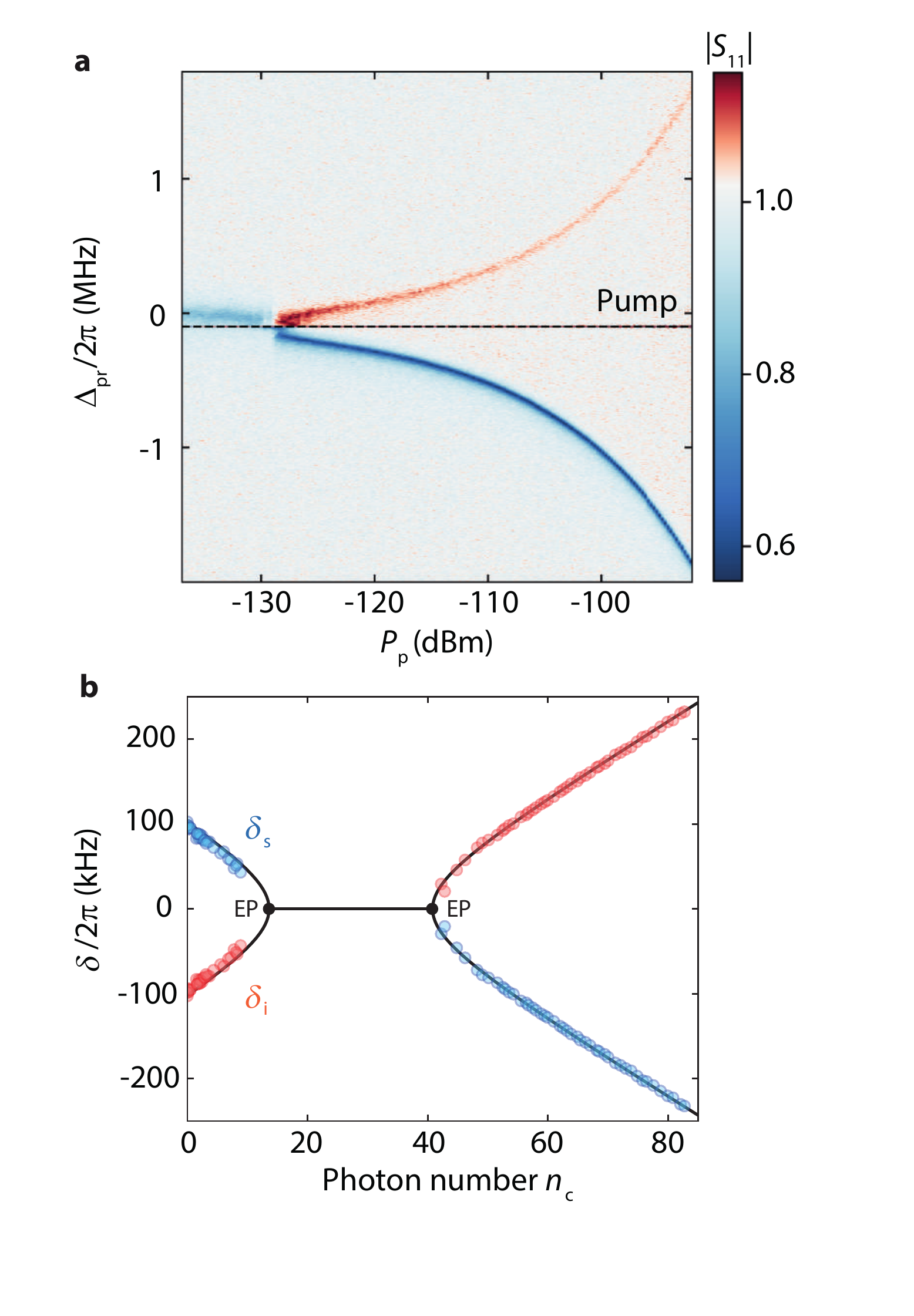}
\caption{\textbf{Level attraction between the signal and idler modes}. Panel \textbf{a} shows the response spectrum magnitude $|S_\textrm{11}|$ versus probe detuning $\Delta_\textrm{pr}$ and on-chip pump power $P_\textrm{p}$. The dashed horizontal line identifies the frequency of the pump, which is placed slightly below the cavity bare resonance frequency with $\Delta_\textrm{p}/2\pi = -100$~kHz. For every linescan we perform a fit of the full response and extract the signal and idler resonance frequencies. Simultaneously, for every linecut we estimate the intracavity photon number $n_\textrm{c}$ based on a fit to the normalized response using the full Kerr model (see Supplementary Note 3). In panel \textbf{b} we plot the extracted signal ($\delta_s$) and idler ($\delta_i$) resonance frequency differences from the pump frequency $\omega_\textrm{p}$ versus $n_\textrm{c}$ as blue and red circles, respectively. The black line is a theory curve based on Eq.~(\ref{eq:signalidler}). We observe level attraction between the two modes and the emergence of two exceptional points (EP). This pinpoints the appearance of an instability region where the signal and idler overlap in frequency. A regime which is characterized by the emergence of an additional imaginary component in the complex resonant solutions $\tilde{\omega}_\textrm{i/s}$, meaning that in this region the modes will acquire different linewidths. Furthermore, we do not show any experimental data in the instability region as the presence of the drive and the fast switching of the cavity between the two sides enclosing the instability region imposes several challenges in the fitting routine.}
\label{fig5:levelattraction}
\end{figure}
At last, {we} quantitatively investigate the agreement between theory and experiment by extracting the magnitude of the response spectrum $|S_\textrm{11}|$ on resonance with the signal and idler modes, and the cavity decay rate. These are compared with the analytically estimated values in Fig.~\ref{fig4:wt_sim}\textbf{c} and \textbf{d}, respectively. While the magnitude of the signal mode $|S_\textrm{11} (\omega_\textrm{s})|$ (full black line) is decreased as we initially reduce the pump detuning, the idler magnitude $|S_\textrm{11} (\omega_\textrm{i})|$ (full red line) is increased further above the background, reaching a maximum output gain of $\sim2$~dB.
We note that in this region the system still operates in a single solution regime. As shown in Fig.~\ref{fig2:blue_detune}, the total cavity linewidth should be dependent on the intracavity photon number, thereby decreasing as we reach the bifurcation point. Past this limit, two new branches emerge. From our theoretical calculations we see that the cavity remains in the high-amplitude branch for just a small detuning range after crossing its bifurcation threshold and shortly after it switches to the low-amplitude branch where the idler mode is afterwards undetectable. For more details on additional experimental data where we achieve $\sim 8$~ dB gain of output gain at the idler resonance see in Supplementary Note 6.\\


The transition between the two amplitude branches of a driven nonlinear system might appear abrupt and experimentally difficult to predict. Nevertheless, carefully exploring this transition regime can unravel interesting phenomena. For this purpose we once again fix the pump on the red side of the cavity, but this time only slightly detuned from its bare resonance frequency with $|\Delta_\textrm{p}| < \kappa$. After this the pump power is slowly increased and the probe strength is kept low to avoid triggering the cavity to prematurely switch. A colormap of the cavity response $|S_\textrm{11}|$ versus probe detuning $\Delta_\textrm{pr}$ and on-chip pump power $P_\textrm{p}$ is shown in Fig.~\ref{fig5:levelattraction}\textbf{a}. Whereas the cavity is initially attracted to the drive tone, once the power reaches a certain threshold ($P_\textrm{p} \sim -129$~dBm) the signal mode jumps onto the other side of the pump and the idler mode simultaneously emerges. By further increasing the power, the modes are pushed further away from the pump and continue resonating symmetrically with respect to the drive tone with complex resonant solutions
\begin{equation}
\tilde{\omega}_\textrm{i/s} = \omega_\mathrm{p} + i\frac{\kappa}{2} \pm \sqrt{\left( \Delta_\mathrm{p} - \mathcal{K}n_\mathrm{c} \right) \left( \Delta_\mathrm{p} - 3\mathcal{K}n_\mathrm{c} \right)},
\label{eq:signalidler}
\end{equation}
where the resonance frequency of the idler and signal modes are given by $\omega_\textrm{i/s} = \textrm{Re}~(\tilde{\omega}_\textrm{i/s})$ and their linewidths by $\kappa_\textrm{i/s} = 2\textrm{Im}~(\tilde{\omega}_\textrm{i/s})$.

We fit each of the linescans of Fig.~\ref{fig5:levelattraction}\textbf{a} with a double-mode reflection response function (see Supplementary Note 2) and for each of them extract the resonance frequency of each of the modes. We note that due to the small amplitude of the idler mode in the linescans below the jump, its experimental resonance frequency remains unfortunately undetected. However for completeness, in this region we plot the idler resonances as mirrors of the signal mode as one would expect from Eq.~(\ref{eq:signalidler}).
The frequency difference between the signal/idler modes and the pump $\delta_\textrm{s/i} = \omega_\textrm{s/i} - \omega_\textrm{p}$ is plotted versus intracavity photon number in Fig.~\ref{fig5:levelattraction}\textbf{b}. 
Besides clearly depicting the frequency symmetry between the two modes, which had already been evident in previously presented data, Fig.~\ref{fig5:levelattraction}\textbf{b} also unveils a very interesting regime where both modes overlap in frequency. This has been observed in other systems \cite{Bernier14,Eleuch14,Seyranian05,Bernier18,Xu16} and it has been refereed to as level attraction \cite{Bernier18}. The exceptional point (EP) where their resonance frequencies meet marks the beginning of a instability regime in which the resonant solutions $\tilde{\omega}_\textrm{i/s}$ will develop an additional imaginary component. This leads to the modes acquiring an inverse modification to their bare linewidth. In Supplementary Note 4 we show the theoretical predictions for the change in linewidth of the modes in the level attraction region. The appearance of such an exceptional point discloses the high susceptibility of a Kerr cavity to noise inducing fluctuations in its intracavity photon number, as in this regime the presence of noise can have a strong impact on its resonance frequency. This is more critical when the cavity is operated closer to the instability regime, as the bending near to the exceptional point is more accentuated (see Supplementary Note 5 for more details).

\section*{Discussion}
We have conducted an experimental study of a Josephson cavity operating in the strong-driving limit and accurately modelled its two-tone response by means of the linearized equation of motion of a Kerr cavity, including its power dependent losses. 
Besides modelling the cavity dispersive shift and its TLS losses, we have detected the emergence of an idler mode upon approaching and crossing the bifurcation threshold of the system. 
We provide an intuitive physical picture for the origin of this mode and present quantitative predictions for its amplitude. 
The appearance of an idler resonance allows us to amplify signals far outside of the cavity lineshape, with the advantage that the interference between the probe and the amplified signals is not affected by the cavity response. 
While most applications of parametric amplifiers operate their systems below their bifurcation point, here we show that for the case of a high-$Q$ resonator, operations above the bifurcation threshold do not necessarily compromise the maximum achievable output gain. 
%


%
Furthermore, in this work we have uncovered an interesting phenomenon where the system hits an exceptional point and there is a level attraction between the eigenfrequencies of the driven resonator.
While this has been studied in other systems, it has only now been observed between the signal and idler modes of a weakly damped nonlinear circuit. 
In the context of Kerr optomechanics, understanding the appearance of this level attraction could be extremely valuable for engineering systems that maximize the driving power of a red-sideband cooling tone.
In addition to the expanded possibilities regarding the manipulation of Josephson circuits beyond their bifurcation threshold, our results ultimately allow for topological control of strongly driven Kerr resonators by encircling its exceptional points upon the implementation of a tunable two-dimensional parameter space \cite{Bernier18, Xu16}.

\subsection*{Acknowledgements}
\vspace{-2mm}

This research was supported by the Netherlands Organisation for Scientific Research (NWO) in the Innovational Research Incentives Scheme -- VIDI, project 680-47-526.
This project has received funding from the European Research Council (ERC) under the European Union's Horizon 2020 research and innovation programme (grant agreement No 681476 - QOMD) and from the European Union's Horizon 2020 research and innovation programme under grant agreement No 732894 - HOT.

\section*{Author contributions}

F.F.S. conceived and conducted the experiments. D.B. and I.C.R. designed and fabricated the device. F.F.S. and I.C.R. performed the data analysis and prepared the first draft of the manuscript. D.B. formulated the theoretical model. All authors discussed the data and the manuscript. G.A.S. supervised the project.

\subsection*{Competing interest}
\vspace{-2mm}
The authors declare no competing interests.

\section*{References}
\bibliography{Refs}

\clearpage{}
\newpage{}

\renewcommand{\thefigure}{S\arabic{figure}}
\renewcommand{\theequation}{S\arabic{equation}}
\renewcommand{\thesection}{S\arabic{section}}
\setcounter{figure}{0}
\setcounter{equation}{0}

\section*{Supplementary Material}

\subsection*{Supplementary Note 1: Device and setup}
\label{subsection:Device_Setup}
The device studied here is the SQUID cavity belonging to the photon-pressure system studied in Ref.~\cite{Bothner21PP}. The cavity was patterned by means of Electron Beam Lithography (EBL) and subsequently loaded into a sputtering machine where a 20 nm layer of Aluminum was deposited. After the deposition, the chip was placed in the bottom of a beaker containing a small amount of anisole and inserted in a ultrasonic bath for a few minutes. After this fabrication process, we proceeded with the patterning of the remaining structures of the photon-pressure system. A step-by-step description of the fabrication is given in the Supplementary Material of Ref.~\cite{Bothner21PP}.

All the experiments reported here were performed in a dilution refrigerator operating at a base temperature close to $T_\textrm{b} = 15$~mK. The printed circuit board (PCB), onto which the sample was glued and wirebonded, was placed in a radiation tight copper housing and connected to a coaxial line used as input/output port. As the device was measured in a reflection geometry, the input and output signals were split via a directional coupler on the $15$~mK stage. The output signal subsequently went into a cryogenic HEMT (High Electron Mobility Transistor) amplifier operating between 4-8 GHz. 


Outside of the fridge we used two different setup configurations. For the single-tone measurements we used a Vector Network Analyzer to characterize the reflection parameter $S_{11}$. For the two-tone spectroscopy we in addition utilized a signal generator as pump tone. The tone subsequently went through a high pass filter and was afterwards combined with the probe signal coming from the VNA by means of a power combiner. Note that the data presented in Supplementary Note 6 was taken in an earlier cooldown and there the pump tone was combined with the probe by means of a room-temperature directional coupler, where the probe was attached to the coupled port and therefore attenuated by 10 dB.

\subsection*{Supplementary Note 2: Fitting routine}
\label{subsection:Fitting}
\subsection*{A. Single-mode response spectrum $S_{11}$}
The $S_{11}$ response function of a parallel LC circuit capacitively coupled to a transmission line in a reflection geometry is given by
\begin{equation}
S_{11} = 1 - \frac{2\kappa_\mathrm{e}}{\kappa_\mathrm{i}+\kappa_\mathrm{e}+2i\Delta}
\label{eq:ResponsefuncHF}
\end{equation}
with detuning from the resonance frequency
%
\begin{eqnarray}
\Delta = \omega - \omega_0.
\end{eqnarray}

We used this expression to model the single-mode linescans of the data presented in main paper.

\subsection*{B. Double-mode response spectrum $S_{11}$}

When the cavity is driven close/beyond its bifurcation point we witness the appearance of an idler mode. In this regime the modes resonance frequencies can be extracted by fitting the full response spectrum with a double resonance response function given by 
\begin{equation}
S_{11} = 1 - \frac{2\kappa_\mathrm{e}^\textrm{s}}{\kappa+2i\Delta_\textrm{s}} - \frac{2\kappa_\mathrm{e}^\textrm{i}}{\kappa+2i\Delta_\textrm{i}}
\label{eq:doubleS11}
\end{equation}
Where $\kappa_\mathrm{e}^{s}, \kappa_\mathrm{e}^{i}$ are the effective external decay rates of each mode, and where the detuning from the signal and idler resonance frequencies are given by

\begin{eqnarray}
\Delta_\textrm{s} = \omega - \omega_\textrm{s}\\
\Delta_\textrm{i} = \omega - \omega_\textrm{i}
\end{eqnarray}

\subsection*{C. Real response spectrum $S_{11}$ and fitting routine}

When analyzing the measured cavity response, we consider a frequency-dependent complex-valued reflection background with amplitude and phase modulations originating from a variety of microwave components in our input and output lines and possible interfering signal paths
\begin{eqnarray}
S_{11}^\textrm{back} = (\alpha_0 + \alpha_1\omega)e^{i(\beta_1\omega + \beta_0)}
\label{eqn:Fit_BG}
\end{eqnarray}
Taking into account the impact of this background on the ideal response functions, the measured response spectrum is given by 
\begin{eqnarray}
S_{11}^\textrm{meas} = (\alpha_0 + \alpha_1\omega) S_{11}e^{i(\beta_1\omega + \beta_0)}
\end{eqnarray}
where we include an additional rotation of the resonance circle with the phase factor $e^{i\theta}$.
The first step in the fitting routine removes the cavity resonance part from the data curve and fits the remaining background with Eq.~(\ref{eqn:Fit_BG}).
After removing the background contribution from the full dataset by complex division, the resonator response is fitted using the ideal response function Eq.~(\ref{eq:ResponsefuncHF}) for the single mode data.
In the final step, the full function is re-fitted to the bare data using as starting parameters the individually obtained fit numbers from the first two steps.
From this final fit, we extract the final background fit parameters and remove the background of the full dataset by complex division.
We note that for the data where both signal and idler modes are present, we calculate the background based on a fit from a single mode linescan and afterwards divide it off from the data. Posterior to this procedure, the double-mode normalized data is fitted with Eq.~(\ref{eq:doubleS11}) for the extraction of the resonance frequencies and with Eq.~(\ref{eq:S11}) for the extraction of the intracavity photon number.
For the data presented in Fig.~\ref{fig4:wt_sim}, as we do not have access to the phase information of the signal due to the measurement scheme used for the acquisition of the data, we exceptionally resorted to a magnitude fitting routine where the background offset was included in the fit model.

\subsection*{Supplementary Note 3: Theory of a driven Josephson cavity}
\label{subsection:Theory}
\subsection{A. Equation of motion}
We model the classical intracavity field $\alpha$ of the Josephson cavity using the equation of motion
\begin{equation}
\dot{\alpha} = \left[i(\omega_0 + \mathcal{K}|\alpha|^2) - \frac{\kappa}{2}\right]\alpha + i\sqrt{\kappa_\mathrm{e}}S_\mathrm{in}
\end{equation}
where $\omega_0$ is the bare cavity resonance frequency, $\mathcal{K}$ is the Kerr nonlinearity, $\kappa$ is the total linewidth, $\kappa_\mathrm{e}$ is the external linewidth and $S_\mathrm{in}$ is the input field.
The intracavity field is normalized such that $|\alpha|^2 = n_\mathrm{c}$ corresponds to the intracavity photon number $n_\mathrm{c}$ and $|S_\mathrm{in}|^2$ to the input photon flux (photons per second).
\subsection*{B. Single-tone response}
With a single tone drive field $S_\mathrm{in} = S_\mathrm{p}e^{i(\omega_\mathrm{p}t + \phi_\mathrm{p})}$ and the Ansatz $\alpha_\mathrm{p}e^{i\omega_\mathrm{p}t}$, where $S_\mathrm{p}$ and $\alpha_\mathrm{p}$ are chosen to be real-valued, we get
\begin{equation}
\left[\frac{\kappa}{2}+i\Delta_\mathrm{d}\right]\alpha_\mathrm{p} - i\mathcal{K}\alpha_\mathrm{d}^3 = i\sqrt{\kappa_\mathrm{e}}S_\mathrm{p}e^{i\phi_\mathrm{p}}
\end{equation}
with $\Delta_\mathrm{p} = \omega_\mathrm{p} - \omega_0$ the detuning between the drive and the undriven cavity resonance frequency.
From this, by multiplication with its complex conjugate, we obtain a third order polynomial for the intracavity photon number $n_\mathrm{c} = \alpha_\mathrm{p}^2$, which is given by
\begin{equation}
\mathcal{K}^2n_\mathrm{c}^3 - 2\mathcal{K}\Delta_\mathrm{p} n_\mathrm{c}^2 + \left(\Delta_\mathrm{p}^2 + \frac{\kappa^2}{4}\right)n_\mathrm{c} - \kappa_\mathrm{e}S_\mathrm{p}^2 = 0.
\label{eq:poli}
\end{equation}
To obtain the full, complex intracavity field with respect to the drive field, we also need the phase $\phi_\mathrm{p}$, which is given by
\begin{equation}
\phi_\mathrm{p} = \atan2\left(-\frac{\kappa}{2}, \Delta_\mathrm{p} - \mathcal{K} n_\mathrm{c}\right).
\end{equation}
The intracavity field is then given by $\alpha = \sqrt{n_\mathrm{c}}e^{-i\phi_\mathrm{p}}$ and the cavity response spectrum by $S_{11} = 1 + i\sqrt{\kappa_\textrm{e}}\frac{\alpha}{S_\textrm{p}}$.

\subsection*{C. Data on single-tone power dependence}

\begin{figure}[h!]
\centering
\includegraphics[trim = {0.5cm, 0.2cm, 0cm, 1cm}, clip=True,scale=0.75]{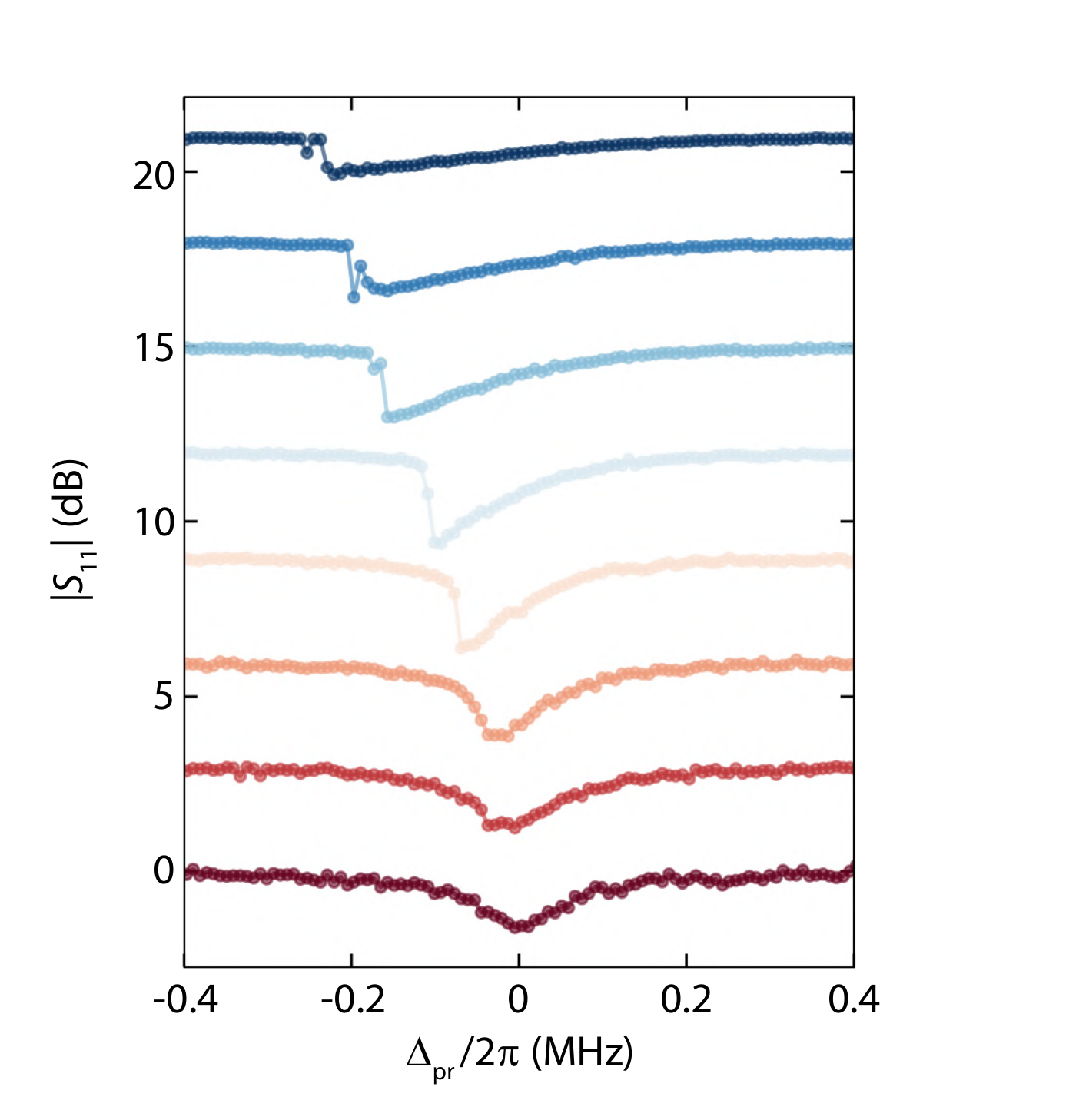}
\caption{\textbf{Observation of the cavity Duffing response through single-tone spectroscopy.} The background corrected cavity response magnitude $|S_{11}|$ is plotted for different values of on-chip probe power. The power used in the first curve is $P_\textrm{pr} = -141$~dBm and this is increased in steps of $2$~dB in the subsequent curves. Furthermore, each of the linescans is up-shifted by $3$~dB for clarity. More details are given in the text.}
\label{figSM2:power_single}
\end{figure}
In Fig.~\ref{figSM2:power_single} we show the single-tone response of the Josephson cavity for increased values of VNA probe power. While for low on-chip powers the cavity exhibits a linear response, i.e. an absorption dip with a Lorentzian lineshape, for higher powers ($P_\textrm{pr}\sim -137$~dBm) the cavity starts bending towards lowers values and we detect a typical Duffing response where there is an abrupt jump from the maximum cavity absorption amplitude to the background level. 
In addition to the bending of the cavity lineshape, which arises from the Kerr non-linearity, the response magnitude on resonance also gets initially deeper. This effect displays the impact of two level system on the cavity internal quality factor. Once the internal decay rate of the cavity reaches its saturation value, the cavity anharmonicity is the only mechanism altering its response and therefore the magnitude on resonance becomes more shallow with power.

Once the cavity crosses its bifurcation point, Eq.~(\ref{eq:poli}) will have three real solutions, where two of them are stable (belonging to the high-amplitude and low-amplitudes branches) and a third unstable solution. Since we scan the probe from lower to higher frequencies, the cavity is mostly operated in the low-amplitude branch. Although for the highest powers we curiously do observe some switching between the branches, which is most probably triggered by noise in the system.

\subsection*{C. Two-tone response}
If the Josephson cavity is driven by a strong drive field and a weaker second input field at frequency $\omega_\mathrm{pr}$, we write for the total input field
\begin{equation}
S_\mathrm{in} = S_\mathrm{p}e^{i(\omega_\mathrm{p}t + \phi_\mathrm{p})} + S_\mathrm{pr}e^{i\omega_\mathrm{pr}t}
\end{equation}
and as Ansatz for the intracavity field we choose
\begin{equation}
\alpha = \alpha_\mathrm{p}e^{i\omega_\mathrm{p}t} + \gamma_-e^{i\omega_\mathrm{pr}t} + \gamma_+e^{i(2\omega_\mathrm{p}-\omega_\mathrm{pr})t}
\end{equation}
with complex-valued amplitudes $\gamma_-$ and $\gamma_+$.
Inserting these into the equation of motion, going to the frame rotating with the signal $\omega_\mathrm{pr}$, linearizing the solution by dropping all terms not linear in $\gamma_-, \gamma_+$ and sorting the equation by frequency components yields three individual equations
\begin{eqnarray}
\left[\frac{\kappa}{2} +i(\Delta_\mathrm{p} - \mathcal{K} n_\mathrm{c})\right]\alpha_\mathrm{p} = i\sqrt{\kappa_\mathrm{e}}S_\mathrm{p}e^{i\phi_\mathrm{p}}\\
\left[\frac{\kappa}{2} +i(\Delta_\mathrm{pr} - 2\mathcal{K} n_\mathrm{c})\right]\gamma_- - i\mathcal{K} n_\mathrm{c}\gamma_+^* =  i\sqrt{\kappa_\mathrm{e}}S_\mathrm{pr}\\
\left[\frac{\kappa}{2} +i(\Delta_\mathrm{pr} - 2\mathcal{K} n_\mathrm{c} + 2\Omega_\mathrm{pp})\right]\gamma_+ - i\mathcal{K} n_\mathrm{c}\gamma_-^* =  0.
\end{eqnarray}
where $\Omega_\mathrm{pp} = \omega_\mathrm{p} - \omega_\mathrm{pr}$ and $\Delta_\mathrm{pr} = \omega_\mathrm{pr} - \omega_0$.
The first of these equations is exactly the same as the one we obtained for single-tone driving.
With the procedure described in the previous section, the intracavity field $\alpha_\mathrm{p}$, the intracavity photon number $n_\mathrm{c}$ and the phase $\phi_\mathrm{p}$ can be determined.
Having solved for $n_\mathrm{c}$ allows then to solve also for $\gamma_-$ and $\gamma_+$.
We write the second and third equations as
\begin{eqnarray}
\frac{\gamma_-}{\chi_\mathrm{p}(0)} - i\mathcal{K} n_\mathrm{c}\gamma_+^*& = & i\sqrt{\kappa_\mathrm{e}}S_\mathrm{pr}\\
\frac{\gamma_+}{\chi_\mathrm{p}(2\Omega_\mathrm{pp})} - i\mathcal{K} n_\mathrm{c}\gamma_-^* & = & 0
\end{eqnarray}
where we defined
\begin{equation}
\chi_\mathrm{p}(\Omega) = \frac{1}{\frac{\kappa}{2} + i\left(\Delta_\mathrm{pr} -2\mathcal{K}n_\mathrm{c} + \Omega\right)}.
\end{equation}
We solve for $\gamma_+$ and get by complex conjugation
\begin{equation}
\gamma_+^* = -i\mathcal{K} n_\mathrm{c} \chi_\mathrm{p}^*(2\Omega_\mathrm{pp})\gamma_-
\end{equation}
Inserting this into the equation for $\gamma_-$ gives
\begin{eqnarray}
\gamma_- & = & i\frac{\chi_\mathrm{p}(0)}{1-\mathcal{K}^2 n_\mathrm{c}^2\chi_\mathrm{p}(0)\chi_\mathrm{p}^* (2\Omega_\mathrm{dp})}\sqrt{\kappa_\mathrm{e}}S_\mathrm{pr}\\
& = & i\chi_\mathrm{g}(0) \sqrt{\kappa_\mathrm{e}} S_\mathrm{pr}
\end{eqnarray}
where in the last step we defined
\begin{equation}
\chi_\mathrm{g}(\Omega) = \frac{\chi_\mathrm{p}(\Omega)}{1-\mathcal{K}^2 n_\mathrm{c}^2 \chi_\mathrm{p}(\Omega)\chi_\mathrm{p}^*(\Omega + 2\Omega_\mathrm{pp})}.
\end{equation}
\subsubsection*{Cavity response function}

The reflection of the driven cavity for a single input probe tone is given by
\begin{eqnarray}
S_{11} & = & 1 + i\sqrt{\kappa_\mathrm{e}}\frac{\gamma_-}{S_\mathrm{pr}} \\
& = & 1- \kappa_\mathrm{e}\chi_\mathrm{g}
\label{eqn:FM_Kerr_S11}
\end{eqnarray}

\subsubsection*{Signal and idler modes}

To find the probe tone resonances of the driven Kerr oscillator, we solve for the complex frequencies $\tilde{\omega}_\mathrm{p}$, for which $\chi_\mathrm{g}^{-1} = 0$.
This is equivalent to
\begin{equation}
	1-\mathcal{K}^2 n_\mathrm{c}^2\chi_\mathrm{p}(0)\chi_\mathrm{p}^*(2\tilde{\Omega}_\mathrm{pp}) = 0
\end{equation}
After multiplying out and sorting for terms with $\tilde{\omega}_\mathrm{p}$, we can write down the two complex solutions as
\begin{eqnarray}
	\tilde{\omega}_\textrm{i/s} = \omega_\mathrm{p} + i\frac{\kappa}{2} \pm \sqrt{\left(\Delta_\mathrm{p} - \mathcal{K} n_\mathrm{c}\right)\left(\Delta_\mathrm{p} - 3\mathcal{K} n_\mathrm{c}\right)},
\end{eqnarray}
where the resonance frequency of the idler and signal modes are given by $\omega_\textrm{i/s} = \textrm{Re}~(\tilde{\omega}_\textrm{i/s})$ and their linewidths by $\kappa_\textrm{i/s} = 2\textrm{Im}~(\tilde{\omega}_\textrm{i/s})$.\\

\subsection*{Supplementary Note 4: Level attraction}
\label{subsection:Level_attraction}
\subsection*{A. Modification to the bare decay rates}

As described in the main paper, upon entering the instability regime characteristic of level attraction, i.e. the regime where the signal and idler mode overlap in frequency, the modes also acquire different decay rates. This is captured by the emergence of an additional imaginary component in the complex resonant solutions given by Eq.~(\ref{eq:signalidler}). 

Our major experimental resource to study this regime is the cavity response $S_{11}$ and within the range enclosed by the two exceptional points we observe that the system displays a single resonance dip in the spectrum. Even though this resonance lineshape is modified due to the frequency overlap of the two modes, as the total response of the system still remains a single resonance dip, for a small difference between their decay rates the result of a standard $S_{11}$ fit becomes inaccurate. Furthermore, as we the cavity and idler modes simultaneously overlap with the pump tone, fitting the cavity response becomes challenging.
Nevertheless, by using the parameters resulting from the fit of the data presented in Fig.~\ref{fig5:levelattraction} we can plot the theoretical estimated values for the change in linewidth of the signal and idler modes. Note that here we did not take into account the power dependent internal losses but only the effects arising from the level attraction.

\begin{figure}[h!]
\centering
\includegraphics[trim = {0.2cm, 0.8cm, 0cm, 0cm}, clip=True,scale=0.6]{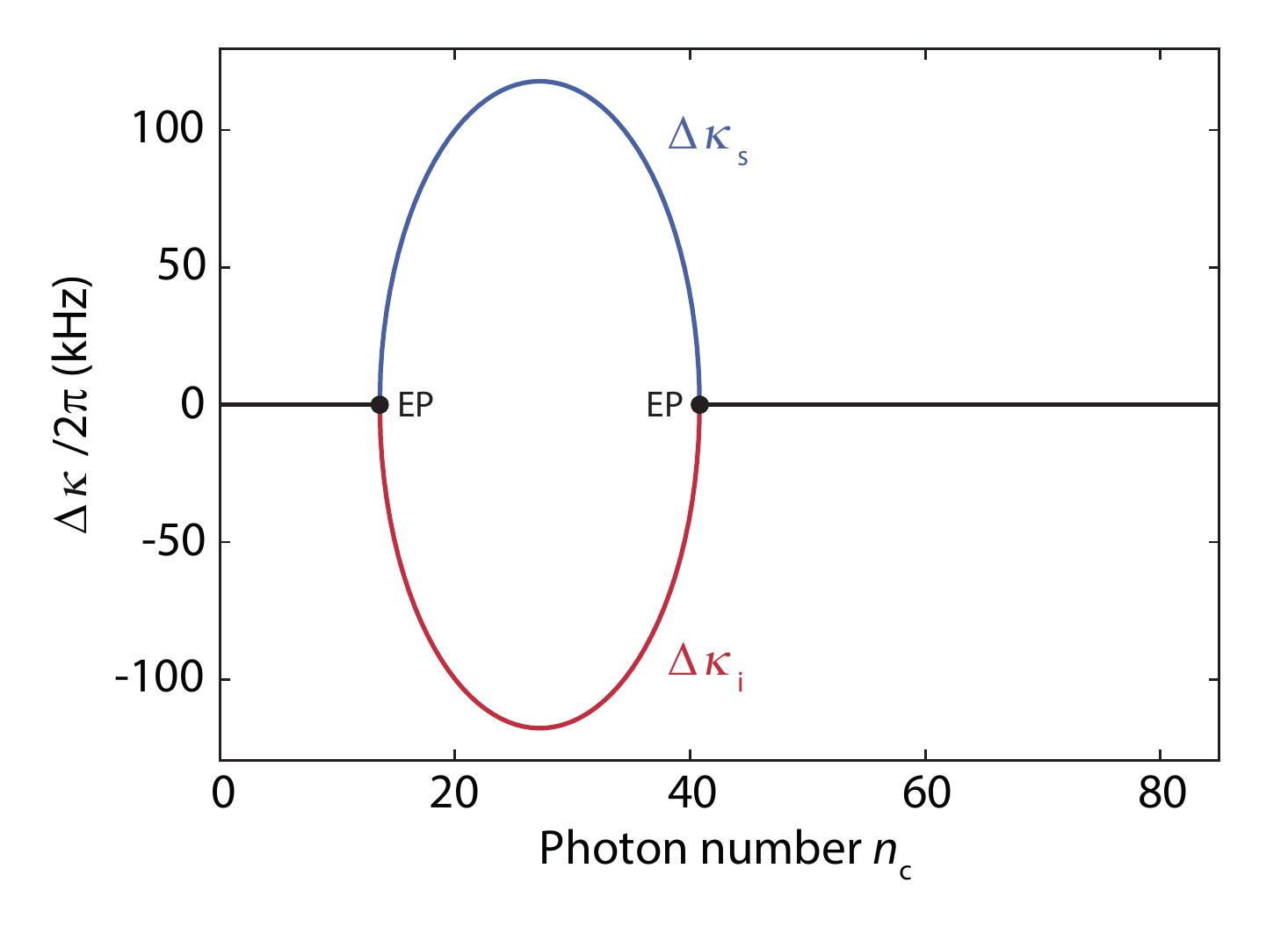}
\caption{\textbf{Modification of the signal and idler decay rates within the regime of level attraction}. The red/blue lines describe the modification to the idler/signal decay rates versus intracavity photon number. The black line describes the region where their linewidths overlap. The lines are the result of analytical calculations for the parameter regime of Fig.~\ref{fig5:levelattraction}. Outside of the region enclosed by the two exceptional points (EP) the signal and idler modes display identical linewidths and distinct resonance frequencies. For the instability region where the modes overlap in frequency, they also acquire different linewidths. While the idler decay rate decreases, the signal decay rate is enhanced. In the operation regime of Fig.~\ref{fig5:levelattraction} of the main paper, the maximum modification to their bare linewidths is predicted to reach $\sim 2\pi\cdot120$~kHz.}
\label{figSM3:linewidths_LA}
\end{figure}

As shown in Fig.~\ref{figSM3:linewidths_LA}, outside of the instability regime, the signal and idler modes should have identical decay rates. However, as the systems enters this range enclosed by the two exceptional points, the idler (red line) should have a reduction in total decay rate while the signal (blue line) should acquire a larger linewidth. For the cavity parameters used in Fig.~\ref{fig5:levelattraction} we estimate that the signal and idler modes should have a maximum modification to their bare decay rates of $\sim 2\pi\cdot120$~kHz.

\subsection*{Supplementary Note 5: Limitations of red-sideband driving in Kerr optomechanics}
\label{subsection:distance_SI}
\begin{figure}[h!]
\centering
\includegraphics[trim = {0cm, 4.7cm, 1cm, 1cm}, clip=True,scale=0.7]{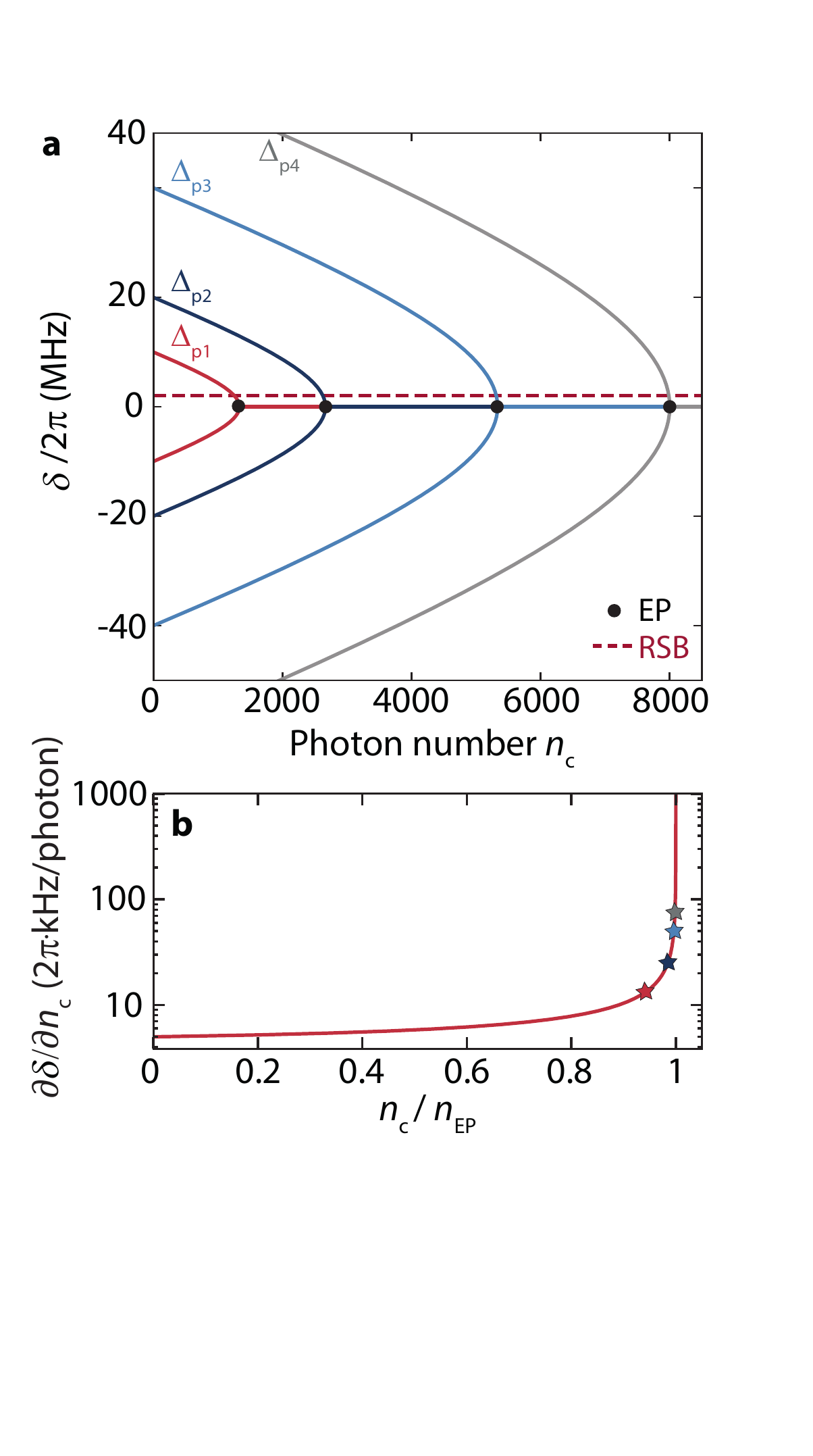}
\caption{\textbf{Level attraction dependence on $\Delta_\textrm{p}$}. In panel \textbf{a} we plot the frequency difference between the cavity/idler mode and the pump, for different values of bare pump detuning $\Delta_\textrm{p} = \omega_\textrm{p} - \omega_0$. Here $\Delta_\textrm{p1} = -2\pi \cdot 10$~MHz, $\Delta_\textrm{p2} = -2\pi \cdot 20$~MHz, $\Delta_\textrm{p3} = -2\pi \cdot 40$~MHz and $\Delta_\textrm{p4} = -2\pi \cdot 60$~MHz. The horizontal red dashed line corresponds to the point where the signal mode and the pump are detuned by one mechanical frequency $\Omega_\textrm{m} = 2\pi \cdot 2$~MHz. The black dots represent the exceptional point for each pump detuning. The intracavity photon number necessary to hit the red-sideband driving condition is $n_\textrm{c} = 1255$, $n_\textrm{c} = 2626$, $n_\textrm{c} = 5313$ and $n_\textrm{c} = 7985$ for $\Delta_\textrm{p1}$, $\Delta_\textrm{p2}$, $\Delta_\textrm{p3}$ and $\Delta_\textrm{p4}$, respectively. In panel \textbf{b} we show the pump photon responsivity of the signal mode versus normalized photon number for each $\Delta_\textrm{p}$, where the normalization factor is the intracavity photon number of their respective exceptional point $n_\textrm{EP}$. As expected, the responsivity versus normalized photon number is independent of $\Delta_\textrm{p}$. Details are given in text.}
\label{SM5:bending}
\end{figure}

A question that arises upon being aware of the instability region characteristic of the level attraction between the signal and the idler modes, is how does this impacts a system which relies on a strong drive field with a fixed frequency distance to the dressed mode. For example, in the field of optomechanics a strong drive tone is crucial to cool mechanical degrees of freedom far below their thermal bath temperatures. As the cooling rate increases with intracavity photon number, the community strives for large pump powers to enhance the multi-photon coupling rates. Furthermore, in cooling protocols the detuning between the cavity and pump is not arbitrary and has to accurately meet the mechanical resonance frequency $\Omega_\textrm{m}$, i.e. $\Delta_\textrm{p} = \omega_\textrm{p} - \omega_0 = -\Omega_\textrm{m}$. The further away the pump is from the red-sideband frequency, the larger is the reduction in the up-scattering rate of pump photons, thereby also reducing the removal of thermal excitations from the mechanical mode. 

In Kerr optomechanics this protocol becomes more challenging as the cavity frequency will shift with power, thereby generating an undesired frequency offset from the cavity red-sideband. 
%
An idea to counter this problem would be to initiate the protocol with a pump tone further detuned from the cavity red-sideband and afterwards increase its power, thereby maximizing the intracavity photon number for the right detuning between the signal mode and the drive.

We explore this possibility in Fig.~\ref{SM5:bending}. Here the frequency difference between the signal/idler mode and the pump is plotted for a variety of pump detunings. Note that in the case of a Kerr cavity the pump detuning $\Delta_\textrm{p}$ refers to the frequency difference between the bare cavity resonance and the pump tone.
From a look to Fig.~\ref{SM5:bending}\textbf{a} it might appear that the right detuning between the signal mode and the pump can be achieved for arbitrary values of photon numbers, just by altering the bare detuning $\Delta_\textrm{p}$. However this ideal scenario is disturbed by the appearance of the exceptional point (EP) already described in Fig.~\ref{fig5:levelattraction}. In fact, as we increase the detuning $\Delta_\textrm{p}$, we need to operate the system closer and closer to this exceptional point in order to preserve the red-sideband driving condition, meaning that the signal mode responsivity to changes in the intracavity photon number will be enhanced.

In Fig.~\ref{SM5:bending}\textbf{b} we plot the pump photon responsivity of the signal mode versus normalized photon number. The normalization factor of each of the curve is given by the intracavity photon number of their respective exceptional point, as this one also depends on $\Delta_\textrm{p}$. Fig.~\ref{SM5:bending}\textbf{b} confirms that the signal mode responsivity increases as we approach the exceptional point. Here, the plotted stars correspond to the pump photon responsivity of the signal mode for the four points described in Fig.~\ref{SM5:bending}\textbf{a}, i.e. where $\omega_\textrm{p} - \omega_\textrm{s} = -\Omega_\textrm{m}$. 

From this figure it becomes clear how the signal mode becomes more susceptible to changes in the intracavity photon number in a cooling protocol using larger values of $\Delta_\textrm{p}$. While for an initial detuning between the bare cavity and the pump of $-10$~MHz one photon changes the signal frequency by $10$~kHz, when $\Delta_\textrm{p}/2\pi = -60$~MHz one photon shifts the signal frequency by $75$~kHz, a value in the order of magnitude as typical decay rates of microwave cavities.
Experimentally this can impose several challenges on the cooling of mechanical modes, as the system becomes highly susceptible to fluctuations of the cavity bare resonance frequency, which subsequently leads to a modification of $\Delta_\textrm{p}$ and changes in the intracavity photon number. For certain detunings, these fluctuations can lead to frequency shifts on the order of a linewidth or even trigger the system to switch between the two sides of the instability regime confined by the two exceptional points (cf. Fig.~\ref{fig5:levelattraction}).

\begin{figure}[h!]
\centering
\includegraphics[trim = {0.6cm, 2cm, 0cm, 0.1cm}, clip=True,scale=0.6]{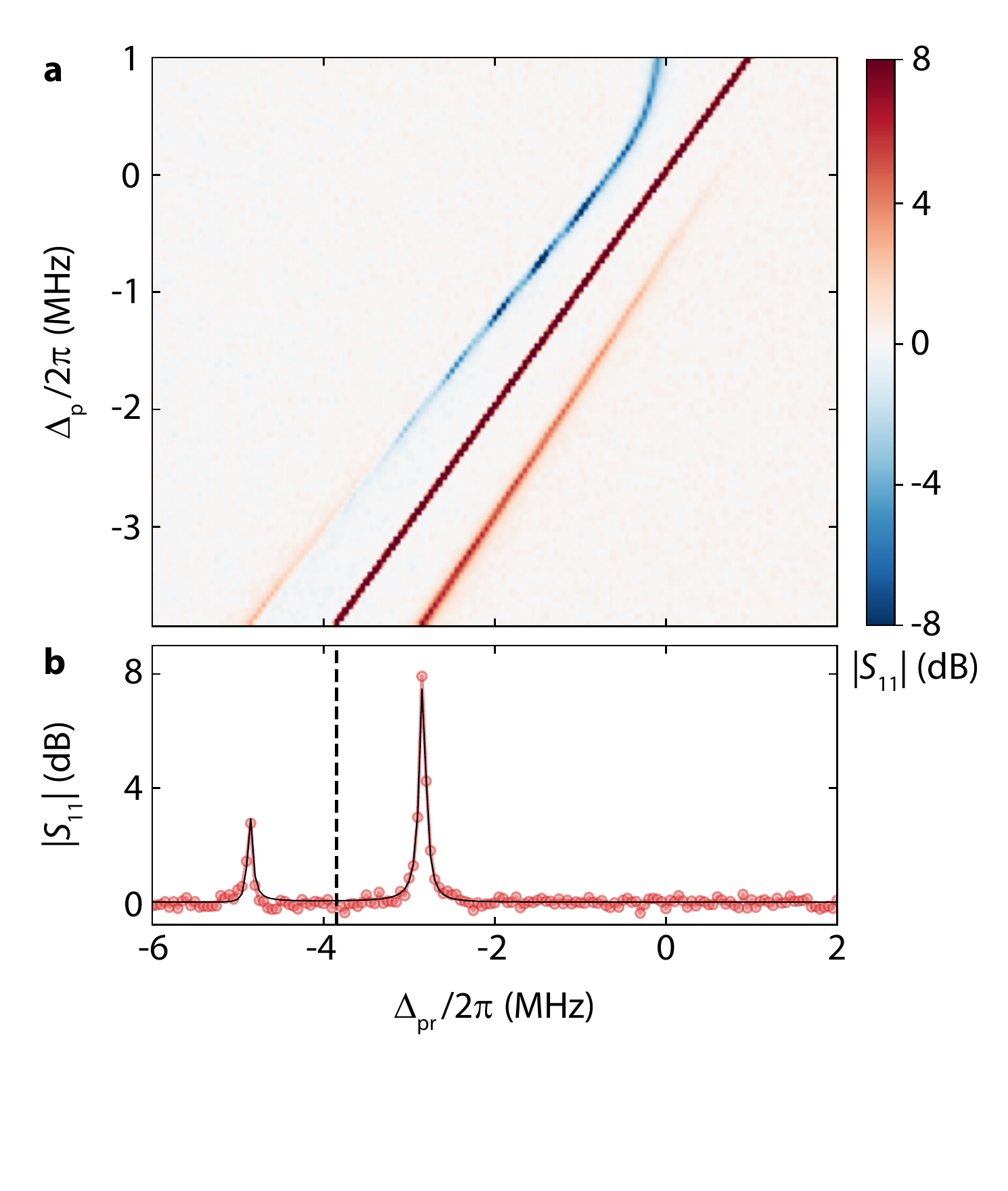}
\caption{\textbf{Observation of net output gain in the signal and idler modes}. Panel \textbf{a} shows a colormap of the normalized cavity response spectrum $|S_{11}|$~(dB) versus probe detuning $\Delta_\textrm{pr}$ and pump detuning $\Delta_\textrm{p}$. The on-chip pump power used here was $P_\textrm{p} = -147 $~dBm. As the pump is swept from higher to lower values the intracavity photon number is enhanced and also the amount of parametric amplification in the device. Note that, contrarily to the data presented in Fig.~\ref{fig4:wt_sim}, here the cavity remains in the high-amplitude branch for the whole range of detunings. In panel \textbf{b} we show a linescan corresponding to the lowest detuning of panel \textbf{a}. Red points are data, the black line is a fit using Eq.(\ref{eq:S11}). The dashed vertial line represents the pump. Here both the signal and idler modes appear as peaks that go above the background level, reveling a regime where the effective internal decay rate of the system is been reduced below zero.}
\label{figSM5:high_gain}
\end{figure}

\subsection*{Supplementary Note 6: Maximizing the cavity output gain}
\label{subsection:highGain}
Fig.~\ref{figSM5:high_gain} shows a similar dataset to the one of Fig.~\ref{fig4:wt_sim}. Here the pump frequency is once again swept from higher to lower values, in such way that it crosses the bare cavity resonance frequency. Fig.~\ref{figSM5:high_gain}\textbf{a} shows the normalized cavity response spectrum $|S_{11}|$ versus probe detuning $\Delta_\textrm{pr}$ and pump detuning $\Delta_\textrm{p}$. In contrast to the data presented in the main paper, in this dataset the cavity remained operating in the high-amplitude branch for the whole range of the measurement. The fact that the cavity did not jump sooner to the low-amplitude branch comes from the fact that we used a smaller probe power compared to Fig.~\ref{fig4:wt_sim}. Furthermore, this data was taken during an earlier cooldown where the device was mounted inside a magnetic shield and possibly there was less noise in measurement setup that could trigger the cavity to switching branches. As the system lingered in this operation regime for a wide range of pump detunings, the intracavity photon number strongly exceeded the maximum amount of drive photons of Fig.~\ref{fig4:wt_sim}, which in that case was achieved immediately prior to the cavity switching point. As shown in Fig.~\ref{figSM5:high_gain}, here not only the idler emerges as a peak above the background, but also the signal mode has turned into a peak. This happens as we reach enough parametric amplification to overcome the internal losses of the cavity. 

Fig.~\ref{figSM5:high_gain}\textbf{b} shows a linescan corresponding to the lowest pump detuning $\Delta_\textrm{p}$ in Fig.~\ref{figSM5:high_gain}\textbf{a}. Here the signal maximum output gain is approximately $3$~dB and the idler reaches a value as high as $8$~dB. This is the largest output gain obtained with this device. This limit could in principle be surpassed in future devices by engineering a higher bare external decay rate.

\end{document}